\documentclass[prb,
superscriptaddress,showpacs,amsmath,amssymb]{revtex4}

\usepackage{graphicx}

\begin{document}

\title{Lifshitz transitions, type-II Dirac and Weyl fermions, event horizon and all that}

\author{G.E.~Volovik}

\affiliation{Low Temperature Laboratory, Aalto University, P.O. Box 15100, FI-00076
Aalto, Finland}

\affiliation{Landau Institute for Theoretical Physics, acad. Semyonov av., 1a,
142432, Chernogolovka, Russia}

\author{K. Zhang}

\affiliation{Low Temperature Laboratory, Aalto University, P.O. Box 15100, FI-00076
Aalto, Finland}

\affiliation{State Key Laboratory of Quantum Optics and Quantum Optics Devices, Institute
of Laser spectroscopy, Shanxi University, Taiyuan 030006, P. R. China}

\date{\today}
\begin{abstract}
The type-II Weyl and type-II Dirac points emerge in semimetals and
also in relativistic systems. In particular, the type-II Weyl fermions
may emerge behind the event horizon of black holes. In this case the
horizon with Painlev\'e-Gullstrand metric serves as the surface of the Lifshitz transition. This relativistic analogy allows us to
simulate the black hole horizon and Hawking radiation using the fermionic superfluid 
with supercritical velocity, and the Dirac and Weyl  semimetals with the interface separating
the type-I and type-II states. The difference between such type of the artificial event horizon and that which arises in acoustic metric is discussed. At the Lifshitz
transition between type-I and type-II fermions the Dirac lines may
also emerge, which are supported by the combined action of topology and
symmetry.
The type-II Weyl and Dirac points also emerge as the intermediate
states of the topological Lifshitz transitions.
Different configurations of the Fermi surfaces, involved in such  Lifshitz transition, are discussed.
In one case the type-II Weyl point connects
the Fermi pockets, and the Lifshitz transition corresponds to the transfer of the Berry flux between the Fermi pockets. In the other case the type-II Weyl point connects the outer and inner Fermi surfaces. At the Lifshitz transition the Weyl point is released from both Fermi surfaces. They loose their Berry flux, which guarantees the global stability, and without the topological support the inner surface disappears after shrinking to a point at the second Lifshitz transition.
These examples reveal the complexity and universality of topological Lifshitz transitions, which  originate from the  ubiquitous interplay of a variety of topological characters of the momentum-space manifolds. For the interacting electrons, the Lifshitz transitions may lead to the formation of the dispersionless  (flat) band with zero energy and singular density of states, which opens the route to room-temperature superconductivity. Originally the idea of the ehancement of $T_c$ due to flat band has been put forward by the nuclear physics community, and this also demonstrates the close connections between different areas of physics.
\end{abstract}
\maketitle

\section{Introduction}
\label{sec:Introduction}

Massless Weyl fermions \cite{Weyl1929} are the building blocks of Standard
Model. In the chiral gauge theory of weak interactions, the fundamental elementary particles are Weyl fermions with a pronounced asymmetry between the $SU(2)$ doublet of left-handed
Weyl fermions and the $SU(2)$ singlet of right-handed Weyl ferrmions. The masslessness of the Weyl fermions is topologically protected.\cite{NielsenNinomiya1981} The corresponding topological invariant
-- the Chern number -- has values $N_{3}=-1$ and $N_{3}=+1$ for the left and right particles respectively.\cite{Volovik2003}   The gapless Weyl fermions are at the origin of the anomalies in quantum field theories, such as chiral anomaly, and the coresponding symmetry protected Chern numbers characterize  the anomalous action.\cite{Volovik2003}
The Dirac particles, which emerge below
the symmetry breaking electroweak transition, are the composite objects obtained by the
doublet-singlet mixing of Weyl fermions with opposite chirality. The topological invariants $N_{3}=\pm 1$ of left and right Weyl fermions cancel each other, and without the topological and symmetry  protection the Dirac particles become massive.

The areal of Weyl fermions is not restricted by the Standard Model of elementary particle physics in general.
Investigations in condensed matter reveal abundant and novel
physics originating from the Weyl fermionic excitations, that live in the vicinity
of the topologically protected touching point of two bands.\cite{NeumannWigner1929,Novikov1981}
Such diabolical (conical) point represents the monopole in the Berry
phase flux,\cite{Simon1983,Volovik1987} and it is described
by the same momentum-space Chern number $N_{3}$.\cite{Volovik2003}
 Weyl fermionic excitations are known to exist in the chiral
superfluid $^{3}$He-A, where the related effects \textendash{} chiral
anomaly\cite{Bevan1997,Volovik2003} and chiral magnetic effect\cite{Krusius1998,Volovik1998}
\textendash{} have been experimentally observed, and in electronic
topological semimetals.\cite{Herring1937,Abrikosov1971,Abrikosov1972,NielsenNinomiya1983,Burkov2011a,Burkov2011b,Weng2015,Huang2015,Lv2015,Xu2015,Lu2015,Hasan2017}
  The Weyl points supported by the higher values of the Chern number, $|N_{3}|>1$, are also possible.\cite{VolovikKonyshev1988} In this case instead of the conical point with linear spectrum of fermions, one has the higher order band touching point, when for example the spectrum is linear in one direction and quadratic in the other directions.\cite{Volovik2003} 
These are the so-called semi-Dirac or semi-Weyl semimetals.\cite{PardoPickett2009,BanerjeePickett2012} 

Recently the attention is attracted to the so-called type-II Weyl
points.\cite{Soluyanov2015,YongXu2015,Chang2016,Autes2016,Xu2016,Jiang2016,Beenakker2016}
A remarkable property of this type of Weyl point is that it is the node of co-dimension 3 in the 3D momentum space, which is
accompanied by the nodes of the co-dimension less than three: 
the nodes of co-dimension 1 (Fermi surfaces) or nodes of co-dimension 2 (Dirac lines).
The transition between the type I and type II Weyl points is the quantum phase transition,   while the symmetry does not necessarily change  at this transition.
The quantum phase transitions with the rearrangement of the topology of the energy spectrum, at which the symmetry remains the same, are called Lifshitz transitions. Originally  I.M. Lifshitz introduced the topological transitions in metals, at which  the connectedness of
the Fermi surface changes.\cite{Lifshitz1959}  Many new types of Lifshitz transition become possible, where the topologically protected nodes of other co-dimensions are involved.\cite{Volovik2017}.
 There is a variety of topological numbers, which characterize the momentum
space manifolds of zeroes. Together with the geometry of the shapes of the manifolds, this makes the Lifshitz transitions widespread in fermionic system.

 In relativistic theories there are several scenarios of emerging of the type-II Weyl points.  
In particular, the transition from the type-I to the type-II Weyl points occurs at the black hole event horizon.\cite{HuhtalaVolovik2002,Volovik2003}
The type II Weyl point may also emerge as the intermediate state of the topological
Lifshitz transition, at which the Fermi surfaces exchange their global topological charge $N_3$.\cite{KlinkhamerVolovik2005a,Volovik2007}
This Weyl point also naturally appears if the relativistic Weyl fermions are not fundamental, but emerge in the
low energy sector of the fermionic quantum vacuum, for example, in the vacuum of the real (Majorana)
fermions.\cite{VolovikZubkov2014} 
These scenarios will be discussed here in connection to the topological materials. Some of
these considerations suggest that the inhomogeneous Weyl semimetal can serve
as a platform for simulating the black hole with stationary
metric and Hawking radiation before the equilibrium is reached. Situations
when the topological invariants are transported between the Fermi surfaces
through type II Weyl point will be considered.

The plan of rest of the paper is as follows. Sec. II describes
the transformation of the type I to type II Weyl fermions
through the intermediate Dirac line. Such transition may occur not only in semimteals, but also
in chiral superfluid, where the transition is regulated by superflow  due to Doppler effect
experienced by Weyl excitations. The symmetry
protected topological number of Dirac line appearing at Lifshithz
transition is discussed. In Sec. III, we consider the behavior of 
the spectrum of Weyl fermions across the event horizon
using the Painlev\'e-Gullstrand space-time. Behind the horizon the Weyl fermions with  type II spectrum
emerge. The Fermi surfaces, which touch each other at the type-II Weyl point, become closed when the Planck scale physics is involved.
Simulation of the event horizon and Hawking 
radiation in Weyl and Dirac semimetals is dicussed. In Sec. IV we consider Lifshitz transitions, which are governed by the interplay of different topological invariants, on example of the transfer of global topological invariants between the Fermi surfaces. In Sec. V the formation of the flat band in the vicinity of the topological transtion is considered. Finally in Sec. VI we review our results and discuss some open questions, in particular in relation to the possibility of room-temperature superconductivity in exotic topological materials.

\section{Dirac line at the transition between type-I and type-II Weyl points}
\label{I-II-transition}

A particular example of emergence of the type-II Weyl fermions in relativistic theories is when
the relativistic Weyl fermions are not fundamental, but represent the fermionic excitations in the
low energy sector of the fermionic quantum vacuum.\cite{FrogNielBook,Volovik2003,Horava2005} The type-I and type-II Weyl fermions may emerge, for example, in the vacuum of the real (Majorana)
fermions.\cite{VolovikZubkov2014} The general form of the relativistic Hamiltonian for the emergent Weyl fermions is obtained by the linear expansion in the vicinity of the topologically protected Weyl
point ${\bf p}^{(0)}$ with Chern number $N_3=\pm 1$: 
\begin{equation}
H=e_{k}^{j}(p_{j}-p_{j}^{(0)})\hat{\sigma}^{k}+e_{0}^{j}(p_{j}-p_{j}^{(0)})\,.\label{HamiltonianGeneral}
\end{equation}
This expansion suggests that the position ${\bf p}^{(0)}$ of the Weyl
point, when it depends on coordinates, serves as the $U(1)$ gauge field, ${\bf A}({\bf r},t)\equiv {\bf p}^{(0)}({\bf r},t)$, acting on relativistic fermions. The parameters $e_{k}^{j}({\bf r},t)$
and $e_{0}^{j}({\bf r},t)$ play the role of the emergent tetrad fields, describing the gravity experienced by Weyl fermions. 

The energy spectrum of the Weyl fermions depends on the ratio between
the two terms in the rhs of Eq.(\ref{HamiltonianGeneral} ), i.e.
on the parameter $|e_{0}^{j}[e^{-1}]_{j}^{k}|$.\cite{VolovikZubkov2014}
When $|e_{0}^{j}[e^{-1}]_{j}^{k}|<1$ one has the conventional Weyl point. The Weyl cone is tilted, if $e_{0}^{j}\neq 0$. At $|e_{0}^{j}[e^{-1}]_{j}^{k}|>1$ the cone is overtilted, and
two Fermi surfaces appear, which touch each other at the Weyl point.
In condensed matter this regime is called the type-II Weyl, as distinct from the conventional Weyl point, which is called type-I.\cite{Soluyanov2015} The Lifshitz transition between the two regimes occurs at $|e_{0}^{j}[e^{-1}]_{j}^{k}|=1$. In the relativistic regime, the spectrum of Weyl fermions at the transition contains zeroes of co-dimension 2 -- 
the Dirac line. In general, the existence of the nodal lines requires the special symmetry: they are protected by topology in combination with symmetry. 

There are indications that in some materials the maximum of the superconducting transition temperature occurs just in the vicinity of the Lifshitz transitions (see also Sec. \ref{Sec:FlatBand}). In particular, the enhancement of $T_c$ at the type-I to-
type-II topological transition in Weyl semimetals has been discussed in Ref.\cite{ShapiroShapiro2017}.

\subsection{Relativistic system}
\label{I-II-transition-Relativistic}

To reveal properties of this Lifshits transition,
let us start with considering the topological charge of the nodal
line using a simple choice of the tetrads for the relativistic Weyl fermions in the gravitational field: 
\begin{equation}
H=c{\mbox{\boldmath\ensuremath{\sigma}}}\cdot\hat{{\bf p}}-fcp_{z}\,.
\label{HamiltonianSimple}
\end{equation}
For $f\neq0$ the Weyl cone is tilted, and for $f>1$ the type-II
Weyl point takes place when the titled Weyl cone crosses zero energy.
At the boundary between the two regimes, with $f=1$, the Hamiltonian has the form
\begin{equation}
H=\begin{pmatrix}0 & c(p_x+ ip_y)\\
c(p_x- ip_y) & -2cp_z
\end{pmatrix}\,,\label{eq:22Matrix}
\end{equation}
and the energy spectrum has the
nodal line on the $p_{z}$-axis, i.e.  $E({\bf p}_{\perp}=0,p_z)=0$ for all $p_z$.
We consider several approaches to characterize stability of the nodal Dirac lines in relativistic systems,
which could be extended to condensed matter systems.

In the first approach we take into account that the matrix in Eq.(\ref{eq:22Matrix}) belongs to the class of
the $2n\times2n$ matrices of the type: 
\begin{equation}
H=\begin{pmatrix}0 & B({\bf p})\\
B^{+}({\bf p}) & C({\bf p})
\end{pmatrix}\,,\label{eq:Matrix}
\end{equation}
and the topological properties of the considered nodes in the spectrum are characteristics of this class.
Of course, it is difficult to expect such matrices in real physical
systems, except for the case of $n=1$, which naturally emerges at Lifshitz transition. But
it is instructive to consider the general $n$ case. The determinant
of such matrix is the product of the determinants of matrices $B$
and $B^{+}$: 
\begin{equation}
D(H)=-D(B)D^{*}(B)\,.\label{eq:Determinant}
\end{equation}
The nodal lines \textendash{} zeroes of co-dimension 2 \textendash{}
are zeroes of $D(B)$ and are described by the winding number of the phase
$\Phi$ of the determinant $D(B)=|D(B)|e^{i\Phi}$: 
\begin{equation}
N_{2}=\oint_{C}\frac{dl}{2\pi i}D^{-1}(B)\partial_{l}D(B)={\bf tr}\oint_{C}\frac{dl}{2\pi i}B^{-1}({\bf p})\partial_{l}B({\bf p})\,,\label{eq:N2B}
\end{equation}
where $C$ is the closed loop in momentum space around the line. 
The line in momentum space with the non-zero winding number of the phase $\Phi$ is the momentum-space analog of the vortex line in superfluids and superconductors, which is characterized by the winding number of the phase of the order parameter.

For
the particular case of $2\times2$ matrix in Eq.(\ref{eq:22Matrix}),
where $D(B)=B=c(p_{X}+ip_{y})$, the invariant can be written as 
\begin{equation}
N_{2}={\bf tr}\oint_{C}\frac{dl}{4\pi i}\cdot[\sigma_{z}H_{f=1}^{-1}({\bf p})\partial_{l}H_{f=1}({\bf p})]\,,\label{eq:N2}
\end{equation}
where the Dirac line corresponds to the $p_{z}$-axis. 

The form $(\ref{eq:N2})$ of invariant $N_2$
is somewhat counterintuitive, since the integral of this type represents the true integer-valued invariant only if $\sigma_{z}$ commutes or anticommutes with the Hamiltonian. The latter  does not happen here, nevertheless the integral is stiil integer-valued, which can be shown in a
straightforward way. For $p_{z}=0$ the Hamiltonian anticommutes with
$\sigma_{z}$, and the integral is the well defined topological invariant
with $N_{2}=1$ for any $f$. At $p_{z}\neq0$ the Hamiltonian does
not anticommute with $\sigma_{z}$. However, Eq.(\ref{eq:N2}) remains
integer for the general $p_{z}$ if $f=1$. 

To see that we apply the second approach. Let us consider
$p_{z}$ as parameter and the arbitrary loops around the line ${\bf p}_{\perp}=0$
with fixed $p_{z}$. Taking into account that 
\begin{equation}
H^{-1}(f=1)=\frac{1}{p_{\perp}^{2}}\left(c{\mbox{\boldmath\ensuremath{\sigma}}}\cdot\hat{{\bf p}}+cp_{z}\right)\,,\label{InverseHamiltonian}
\end{equation}
one obtains that the variation of $N_2$ over $p_z$ is zero:
\begin{equation}
\frac{dN_{2}(p_{z})}{dp_{z}}=0\,\,,\,\,N_{2}(p_{z})={\bf tr}\oint_{C(p_{z})}\frac{dl}{4\pi i}\cdot[\sigma_{z}H_{f=1}^{-1}({\bf p}_{\perp},p_{z})\partial_{l}H_{f=1}({\bf p}_{\perp},p_{z})]\,.\label{eq:N2derivative}
\end{equation}
Thus the integral $N_{2}(p_{z})=1$ for any $p_{z}$ at $f=1$.

Finally,  the stability of the vortex line in momentum space can be understood through consideration in
terms of the determinant of the Hamiltonian matrix, i.e. $D(H)$,
in a way somewhat similar to that in Ref.\cite{KimmeHyart2016} 
\begin{equation}
D(H)=c^{2}p_{z}^{2}(f^{2}-1)-c^{2}p_{\perp}^{2}\,.\label{eq:Drel}
\end{equation}
$D(H)$ is nonzero for $0<f<1$, is zero on line at $f=1$,
and has zeros on the conical Fermi surface at $f>1$. For $f=1$ one
can define the generalized root $q(H)$ of det $H$ \textendash{}
a polynomial function of the matrix elements of the Hamiltonian \textendash{}
in such a way that $|q(H)|^{2}=|D(H)|$. So $q(H)$ is our $D(B)$
in Eq.(\ref{eq:Determinant}). The corresponding polinomial is 
\begin{equation}
q(H_{f=1})=D(B)=c(p_{x}+ip_{y})\,.\label{eq:qH}
\end{equation}
It has zero on the line ${\bf p}_{\perp}=0$ which is protected by
$2\pi$ winding of the phase of $q$ around the line. This gives rise
to the topologically stable zero in the determinant $D(H)$ and thus to
topologically stable zero in the quasiparticle spectrum. For $f\neq 1$,
the integral Eq.(\ref{eq:N2}) depends on $p_{z}$ and on the radius
of the closed loop $C$.

\begin{figure}
\centering 
\includegraphics
{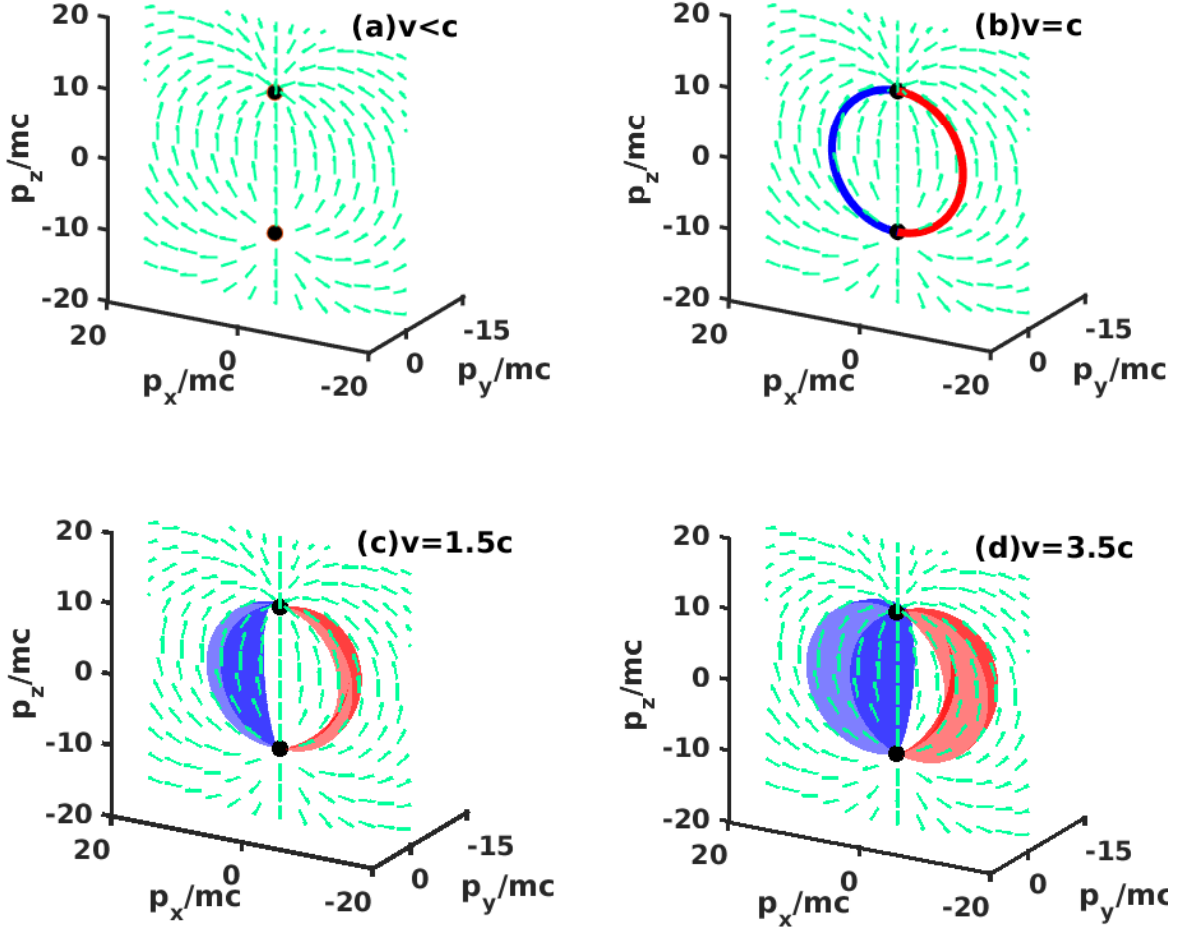} \caption{\textbf{Type-I and type-II Weyl points (black dots) across the Lifshitz transition.} 
When the  superfluid velocity exceeds the pair-breaking
velocity(\char`\"{}speed of light\char`\"{}), the type-I Weyl points in
the original chiral superfluid are converted to the type-II Weyl points. The
process of this Lifshitz transition is shown for Eq.(\ref{Hamiltonian3HeA}) describing quasiparticles in the chiral superfluid $^3$He-A in the presence of superfluid current with velocity
${\bf v}=v\hat{{\bf x}}$. Green arrows depict the vector configurations in the $p_{y}=0$
plane of the momentum space monopole located at Weyl points.  
\\
(\textit{top left}): 
Two original type-I Weyl points at $v<c$.
\\
(\textit{top right}): At the critical speed $v=c$, two Dirac
lines are formed, by which two Weyl points are connected. The
red  and blue lines correspond to the Dirac nodes in the hole 
and particle spectrum of Eq.(\ref{Hamiltonian3HeA}) respectively.
\\
(\textit{bottom}): 
Particle and hole Fermi pockets connected via the type-II Weyl points appear
when $v>c$. }
\label{3HeA} 
\end{figure}


\subsection{Chiral superfluid}
\label{I-II-transition-Chiral}

 It should be pointed out that the Dirac
line emerges for the relativistic fermions mainly due to the Lorentz invariance of the linear spectrum. In principle, the nodal line may disappear when the
higher order nonrelativistic corrections are taken into account, such as
 the Planckian quadratic term of momentum discussed in Sec.\ref{BHhorizon}, if there is no additional symmetry, which could support
the stability of the nodal line.
The nonlinear terms are natural in condensed matter systems, and we consider the energence of the
Dirac line at Lifshitz transition, which can be realized in chiral
superfluid system, such as superfluid $^{3}$He-A. The simple model
Hamiltonian with the Dirac lines existing at Lifshitz transition is:
\begin{equation}
H=p_{x}v+\tau_{3}\frac{p^{2}-p_{F}^{2}}{2m}+\tau_{1}cp_{x}+\tau_{2}cp_{y}\,.
\label{Hamiltonian3HeA}
\end{equation}
Here the Weyl points are in positions $\pm p_F\hat{\bf z}$;  the superfluid velocity with
respect to the heat bath  ${\bf v}=v\hat{{\bf x}}$ is transverse to the direction towards the Weyl points. The first term in the rhs of Eq. (\ref{Hamiltonian3HeA})
comes from the Doppler shift produced by superflow;\cite{Volovik2003} 
$\tau_{i}$ are the Pauli
matrices in the Bogoliubov-Nambu space; in $^{3}$He-A $v_{F}=p_{F}/m\gg c$. Here
$c$ is the maximum speed of quasiparticle propagating in the plane $(p_x,p_y)$ in vicinity of the Weyl points, where the spectrum is relativistic  in the linear expansion of the Hamiltonian. 

The transition between
the type-I Weyl fermions and the type-II Weyl fermions tales place,
when the flow velocity $v$ reaches the \char`\"{}speed of light\char`\"{}
$c$, see Fig. \ref{3HeA}. For $v<c$ there are two Weyl points at
${\bf p}^{(0)}=\pm p_{F}\hat{{\bf z}}$ with opposite topological
charges $N_{3}=\pm1$, and thus with opposite chiralities of the 
relativistic Weyl fermions living near the Weyl points, Fig. \ref{3HeA} (\textit{top left}). 
At $v>c$ there are two banana shape particle and hole Fermi surfaces, 
which contact each other at the type-II Weyl points, see
Fig. \ref{3HeA}  (\textit{bottom}). Exactly at the Lifshitz transition, at
$v=c$, one has Dirac lines, which connect the Weyl points in Fig.
\ref{3HeA} (\textit{top right}). At $v=c$ the matrix Hamiltonian belongs to the class of matrices in Eq.(\ref{eq:Matrix}).
 The corresponding determinant $D(B)$ describing the
topology of the line at $f=1$ in Eq.(\ref{eq:Determinant}) is: 
\begin{equation}
D(B)=\frac{p^{2}-p_{F}^{2}}{2m}+icp_{y}\,.\label{eq:Dnonrel2}
\end{equation}
It has the topologically protected lines of zeroes at $p_{y}=0$,
$p_{x}^{2}+p_{z}^{2}=p_{F}^{2}$ in Fig. \ref{3HeA}(b). These are the vortex lines in momentum space
 with the winding numbers $N_{2}=\pm1$  of the phase of the determinant $D(B)$
  in Eq.(\ref{eq:N2B}), where the contour $C$ is along closed
loop surrounding the Dirac lines.

\subsection{Lifshitz transition with crossing of Dirac lines and Hopf linking}
\label{Crossing}

\begin{figure}
\centering 
\includegraphics
{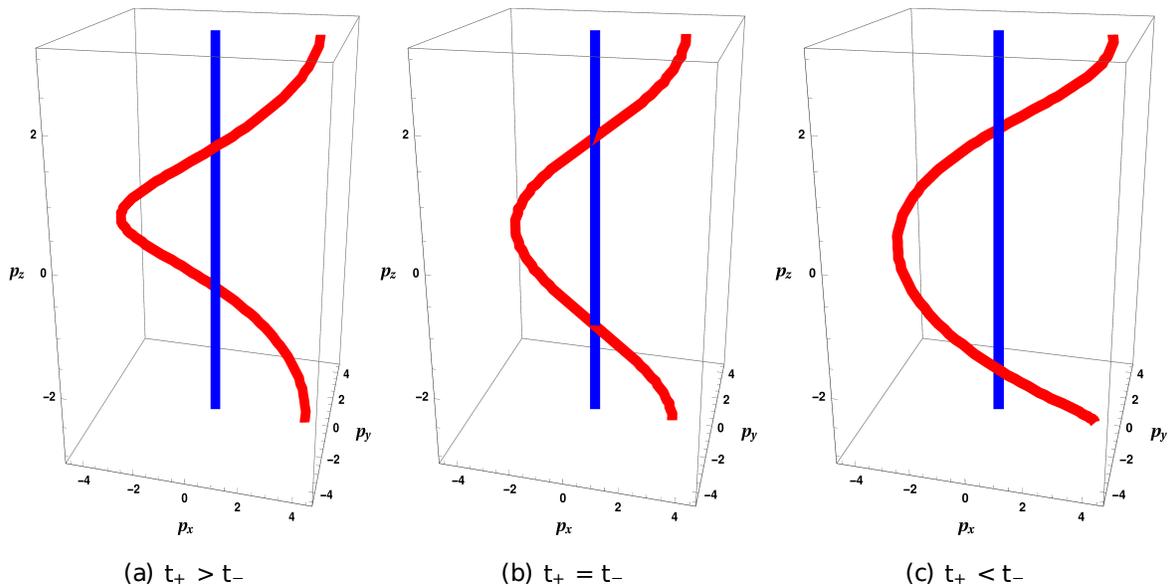} \caption{\textbf{Topological Lifshitz transition of Hopf linked nodal lines.} 
(a) Hopf link with linking number $N_{l}=1$. (b) Lifshitz transition. (c) Hopf link with linking number $N_{l}=-1$.(a) and (c) are not ambient isotopic so that are topologically distinguished.}
\label{Nodals} 
\end{figure}

This type of transtion can be seen on example of the modification of the model describing the rhombohedral graphite
\cite{HeikkilaVolovik2015}, with
\begin{equation}
B= (p_x +ip_y)\left( p_x +ip_y + t_+e^{ip_z a} +  t_-e^{-ip_z a}  \right)\,.
\label{eq:CrossingLines}
\end{equation}
This model has two nodal lines -- the straight one along the $z$ direction and the spiral around the straight one. Due to the lattice periodicity originating from the layers type construction along $z$ direction, the spectrum of quasiparticle along $p_z$ direction can be described in terms of the one dimensional Brillouin zone. As a result the nodes in the spectrum are the closed loops. From the viewpoint of knot theory, these two nodal lines form a Hopf link --  the simplest nontrivial link consisting of two unknots.\cite{Kauffman2001}. The Hopf link with $t_+>t_-$ is the mirror image of the one with $t_+<t_-$, and these two configurations cannot be connected with combination of Reidemeister moves.\cite{Kauffman2001} This means that they are not ambient isotopic, and the may transform to each other only via the special type of Lifshitz transition, which in our case occurs at $t_+=t_-$. To characterize the difference between these kinds of two Hopf linked nodal lines, we assign a fixed direction and calculate the linking number via $N_{l}=(\sum_{p} {\epsilon_{p}} )/2$, where $p$ is the crossing in the diagram of Hopf link of nobal lines, and $\epsilon_{p}$ is the sign of the oriented crossing.\cite{Kauffman2001} From Fig.(\ref{Nodals}), we can find that $N_{l}=1$ for $t_+>t_-$, while $N_{l}=-1$ for $t_+<t_-$ respectively.

On other examples of knotted nodal lines see e.g. in Ref.\cite{Nodal-knot2017}.

\section{Transition between type-I and type-II Dirac/Weyl vacua and event
horizon}
\label{EventHorizon}

The section \ref{I-II-transition-Chiral} demonstrated the scenario of Lifshitz transition between type-I and type-II Weyl points, when the flow velocity of the chiral superfluid liquid exceeds the ``light speed'' of an emergent Weyl quasiparticles.
In Sec. \ref{BHhorizon} we shall see that  analagous transition occurs for the relativistic fermions when the event horizon of the black hole is crossed and the frame drag velocity exceeds the speed of light.  This analogy suggests a route for simulation of an event horizon in inhomogeneous condensed matter systems, which is accompanied by the analog of Hawking radiation. This will be discussed in Sec. \ref{ArtificialBH}.

\subsection{Type-II Weyl fermions behind the black hole horizon}
\label{BHhorizon}

In general relativity the convenient stationary metric for the black hole both
outside and inside the horizon is provided in the Painlev\'e-Gullstrand
spacetime\cite{Painleve} with the line element:
\begin{equation}
ds^{2}=-c^{2}dt^{2}+(d{\bf r}-{\bf v}dt)^{2}=-(c^{2}-v^{2})dt^{2}-2{\bf v}d{\bf r}dt+d{\bf r}^{2}\,.\label{Painleve}
\end{equation}
This is stationary but not static metric, which is expressed in terms
of the velocity field ${\bf v}({\bf r})$ describing the frame drag in the
gravitational field. The Painlev\'e-Gullstrand is equivalent to the so-called acoustic metric, 
\cite{unruh1,unruh2,Kraus1994}
where ${\bf v}({\bf r})$ is the velocity of the normal  or superfluid liquid.

For the spherical black hole the frame drag velocity field (the velocity of the free-falling observer) is radial: 
\begin{equation}
{\bf v}({\bf r})=-\hat{{\bf r}}c\sqrt{\frac{r_{h}}{r}}~,~r_{h}=\frac{2MG}{c^{2}}\,.\label{VelocityField}
\end{equation}
Here $M$ is the mass of the black hole; $r_{h}$ is the radius of
the horizon; $G$ is the Newton gravitational constant. The minus
sign in Eq.(\ref{VelocityField}) gives the metric in case of the black hole,
while the plus sign would characterize the gravity of a white hole. 

Let's us consider a Weyl particle in the Painlev\'e-Gullstrand
space-time. The 
tetrad field corresponding to the metric in Eq.(\ref{Painleve}) has the form:\cite{Doran2000} 
\begin{equation}
e_{k}^{j}=c\delta_{k}^{j}\ \text{and}\ e_{0}^{j}=v^{j}  \,,
\label{PGTetrad}
\end{equation}
which leads to the following Hamitonian:
\begin{equation}
H=\pm c{\mbox{\boldmath\ensuremath{\sigma}}}\cdot{\bf p}-p_{r}v(r)+\frac{c^{2}p^{2}}{E_{{\rm UV}}}\,\,,\,\,v(r)=c\sqrt{\frac{r_{h}}{r}}\,.\label{HamiltonianBH}
\end{equation}
Here the plus and minus signs correspond to the right handed and left
handed fermions respectively; $p_{r}$ is the radial momentum of fermions.
The second term in the rhs of (\ref{HamiltonianBH}) is the Doppler
shift ${\bf p}\cdot{\bf v}({\bf r})$ caused by the frame drag velocity (compare
with Eq.(\ref{Hamiltonian3HeA}) for chiral superfluid). 

The third
term in Eq.(\ref{HamiltonianBH}) is the added nonlinear dispersion to take into account the Planckian physics,
which becomes important inside the horizon. The parameter
$E_{{\rm UV}}$ in the third term is the ultraviolet (UV) energy scale, at which the Lorentz invariance
is violated. The UV scale is typically associated with but does not necessarily correspond to the Planck energy scale.\cite{Kostelecky2011,Rubtsov2016}
 For the interacting fermions such term can arise in effective
Hamiltonian $H=G^{-1}(\omega=0,{\bf p})$ even without violation of
Lorentz invariance on the fundamental level:\cite{Volovik2010} the Green's
function $G(\omega,{\bf p})$ may still be relativistic invariant,
while the Lorentz invariance of the Hamiltonian is violated due to
the existence of the heat bath reference frame. In this case the UV scale is below the Planck scale.

\begin{figure}
\centering
\includegraphics{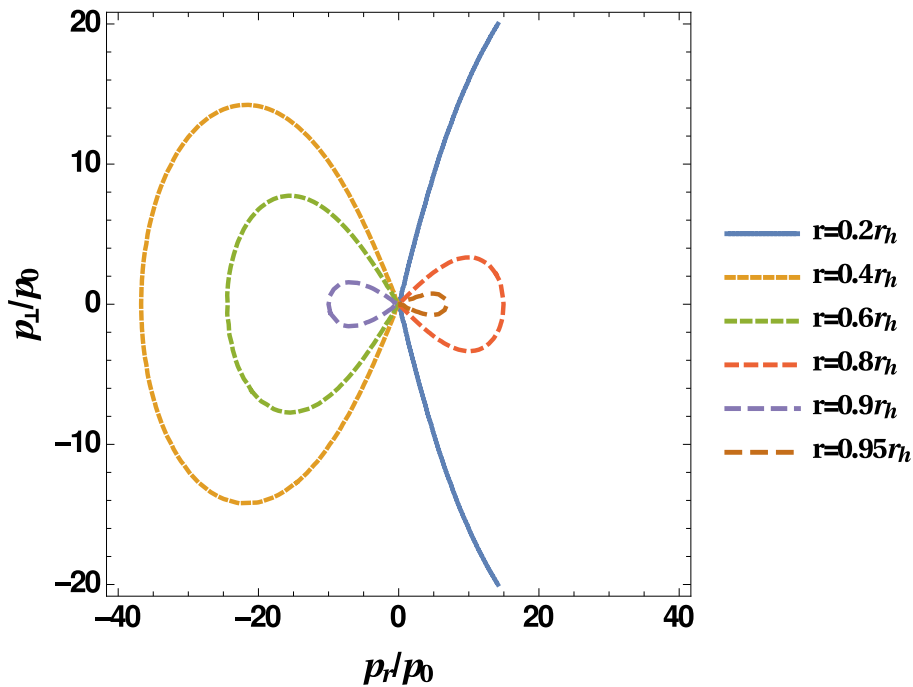} 
\caption{\textbf{Type II Weyl point behind the Black hole event horizon.} Contours
of Fermi surfaces attached to the type-II Weyl fermions of Standard
Model at different radial positions $r$ inside the black hole horizon. Here $p_{r}$ is the radial
component of the momentum ${\bf p}$; $p_{\perp}=\sqrt{p^{2}-p_{r}^{2}}$
and $p_{0}=\hbar/r_{h}$, where $r_h$ is the radius of spherical black holes. The contours with $p_{r}>0$ 
correspond to the Fermi pockets of particles, and those with $p_{r}<0$ are the hole Fermi pockets (Fermi surfaces of anti-particles). For each position $r$ only one of the two Fermi surfaces is shown.
The process of the filling of particle and hole Fermi pockets inside the horizon is observed as the Hawking radiation outside 
the horizon.\cite{Volovik2003}
}
\label{FermiSurfaceBH} 
\end{figure}

In Fig. \ref{FermiSurfaceBH}, we present the Fermi surfaces, which appear behind
the horizon at different positions $r<r_{h}$. Behind the black hole horizon,
the Weyl point in the spectrum transforms to the pair of the closed Fermi
surfaces with the touching point: the type-II Weyl point. The $p^{2}$
term in Eq.(\ref{HamiltonianBH}) makes the Fermi surfaces attached
to the type-II Weyl point closed, while it provides only a small
correction when $cp\ll E_{{\rm UV}}$. The latter is valid 
if the spectrum inside the black hole is considered in the vicinity of the horizon,
where $r_{h}-r\ll r_{h}$, see  Fig. \ref{FermiSurfaceBH},
at positions $r=0.95r_{h}$ and $r=0.9r_{h}$.
 Similar o the situation in section II, near
the horizon the Dirac(Weyl) cone is tilted, and behind the horizon
it crosses zero energy and forms the Fermi surfaces corresponding
to the type-II Weyl points. But there is no  Dirac line at the
horizon, where Lifshitz transition occurs: as we mentioned in section
\ref{I-II-transition-Chiral} this is because of the quadratic term, which violates the Lorentz invariance. 
Instead, the Fermi pockets start to grow from the Weyl point  with $|p_{r}|<p_{{\rm UV}}(v(r)-c)/c\ll p_{{\rm UV}}$,
when the horizon is crossed.

In the full equilibrium the Fermi pockets must be occupied by particles and "holes".
One of the mechanisms of the filling of the Fermi pockets in the process of equilibration will be observed by external observer as Hawking radiation.\cite{Volovik2003} The Hawking temperature is determined by effective gravitational field at the horizon:
\begin{equation}
T_{{\rm H}}=\frac{\hbar}{2\pi}\left(\frac{dv}{dr}\right)_{r=r_{h}}\,
\label{HawkingT}
\end{equation}
If the Hawking radiation is the dominating process of the black hole evaporation, the lifetime of the black hole
is astronomical. However, the other much faster mechanisms involving the trans-Planckian physics are not excluded.

\subsection{Artificial black hole and Hawking radiation from Lifshitz transition}
\label{ArtificialBH}


\begin{figure}[b]
 \includegraphics{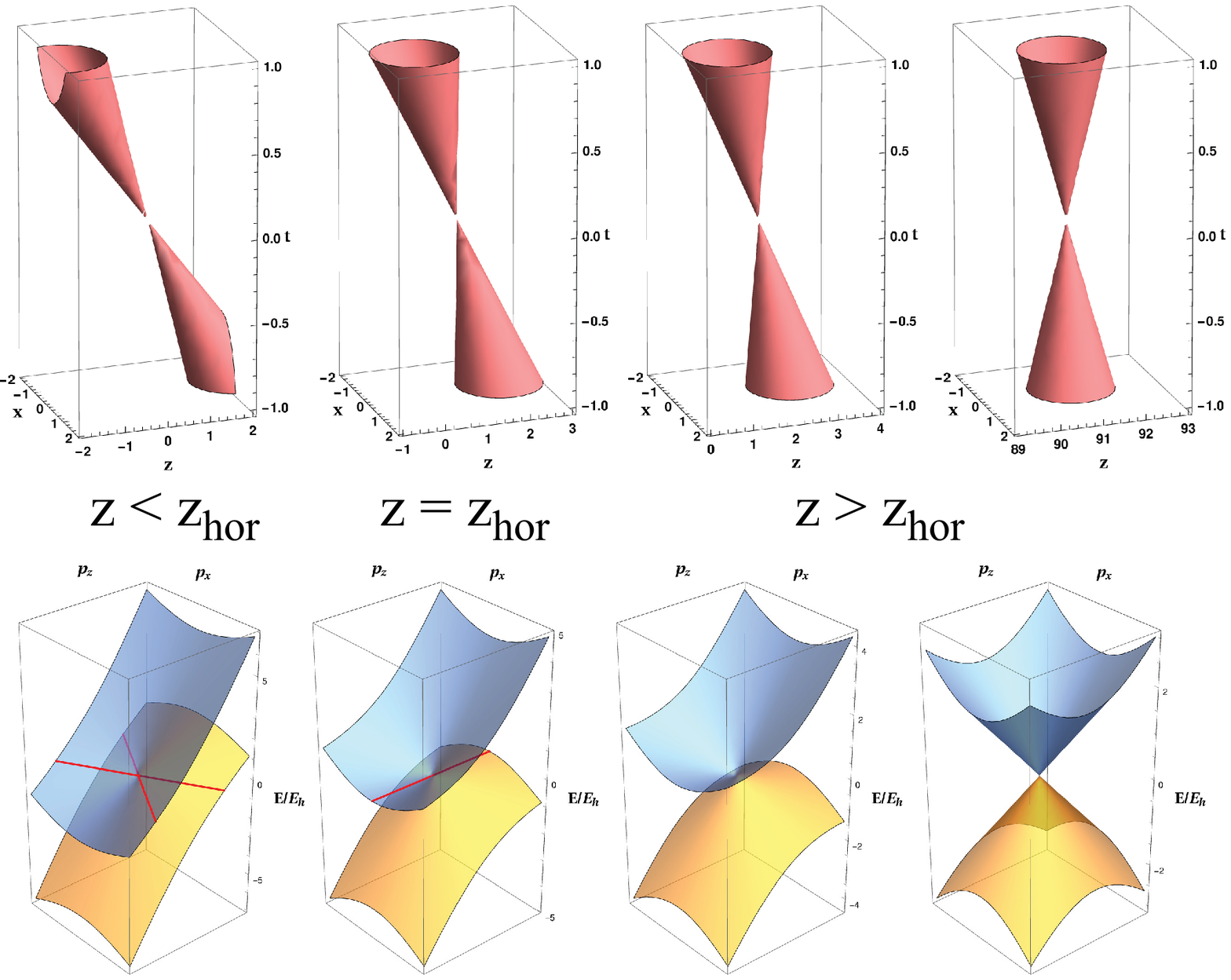} 
\caption{\textbf{Weyl cone and light cone of artificial black hole.} 
The artificial
event horizon can be simulated in Weyl semimetals using the interface between
type-I Weyl material ($z>z_{\rm hor}$) and type-II Weyl material ($z<z_{\rm hor}$). 
With decreasing $z$, the Weyl cones (lower row) and the corresponding  "light cones" for Weyl quasiparticle (upper row) are gradually titled.
In the upper row the light cone is overtilted behind the horizon, so that quasiparticle can move only away from the horizon into the black hole region.
The lower row  demonstrates the process of Lifshitz transition at the horizon. Behind the horizon the Weyl cone is overtilted and two Fermi surfaces appear (red lines correspond to zero energy), connected by type-II
Weyl point.
 Filling of the Fermi surfaces by particles and holes behind the horizon
corresponds to the  Hawking radiation, if the electrons and holes come from the region outside the horizon.
The process of tunneling of quasiparticles from outside the horizon to inside the horizon\cite{Volovik1999,ParikhWilczek}  is seen as Hawking radiation with temperature in Eq.(\ref{HawkingT}) for the black hole or by Eq.(\ref{HawkingTflat}) for the flat horizon.
 After the particle and hole states in the Fermi surfaces are fully occupied,
the Hawking radiation stops.
While in the black hole the shapes of the horizon and ergosurface are determined by Einstein equations,  in semimetals they can be designed.
}
\label{BH} 
\end{figure}


Based on the discussion in Sections \ref{I-II-transition-Chiral}  and \ref{BHhorizon},
one can suggest a new route through which the black hole horizon and ergosurface
can be simulated using the inhomogeneous condensed
systems with emergent type-I and type-II Wely fermions. The interface
which separates the regions of type-I and type-II Weyl points may serve as the event horizon, on which 
the Lifshitz
transition takes place. In general case, such an artificial horizon may
have the shape different from the spherical surface. The shape of the horizon is not important
if we are interested in the local temperature 
of Hawking radiation, which is determined by the local effective gravity at the horizon.

Let us consider the completely flat artificial event horizon on example of Eq.(\ref{HamiltonianSimple}). We assume that the parameter $f$ depends on $z$, and $f(z)$ crosses unity at $z=z_{\rm hor}$.
  The plane $z=z_{\rm hor}$ separates  the region with type-I Weyl fermions ($f(z)<1$) from the region with type-II Weyl 
fermions  ($f(z)>1$).  This plane corresponds to the event horizon, while the ergoplane 
can be obtained for the other orientations of the plane with respect to the axis $z$ (review on
artificial horizons and ergoregions in acoustic metric see in Ref. \onlinecite{Visser1998}). 
Fig. \ref{BH} demonstrates the Weyl cones in the energy-momentum space ({\it bottom}) and the analogues of the light cone ({\it top})  for quasiparticles on two sides of the event horizon.
Behind the horizon the Weyl cone is overtilted so that the upper cone crosses the zero energy level,
and the Fermi surfaces (Fermi pockets) are formed, which are connected by the type-II Weyl point.
Correspondingly the future light cone is overtilted behind the horizon so that all the paths fall further into the black hole region. 

The filling of the originally empty states inside the Fermi surfaces causes the Hawking radiation.
 For the flat horizon the Hawking temperature determined by the effective gravitational field at the horizon is:
\begin{equation}
T_{{\rm H}}=\frac{\hbar c}{2\pi}\left(\frac{df}{dz}\right)_{z=z_{\rm hor}}\,
\label{HawkingTflat}
\end{equation}

Note that in the Weyl semimetals the mechanism of formation of the artificial event horizon (and its behavior
after formation) is different from the traditional mechanism, which is based on the supercritical flow of the liquid or
Bose-superfluids.\cite{unruh1,unruh2,Visser1998,Volovik2003} 
In the latter case the effective metric (the so-called acoustic metric) is produced by the flow of the liquid and
thus represents the non-static state. Due to the dissipation (caused, say, by the analogue of Hawking radiation) the flow relaxes and
reaches the sub-critical level, below which the horizon disappears. On the contrary, in semimetals the tilting of the Weyl cone occurs without
the flow of the electronic liquid, and thus the state with the horizon is fully static. The dissipation after the formation
of the horizon (caused, say, by analogue of Hawking radiation) leads to the filling of the electron and hole Fermi pockets. After the Fermi pockets are fully occupied the final state is reached, but it still contains the event horizon, though the Hawking radiation is absent. Similar
mechanism takes place in the fermionic superfluids, such as superfluid $^3$He, where depending on the parameters of the system the flow may or may not remain  supercritical after the Fermi pockets are occupied, see Fig. 26.1 in Ref.\onlinecite{Volovik2003}.

\section{Lifshitz transitions with type-II Weyl and Dirac points at the transition}
\label{Sec:intermediateWeylDirac}
\begin{figure}
\centering \includegraphics{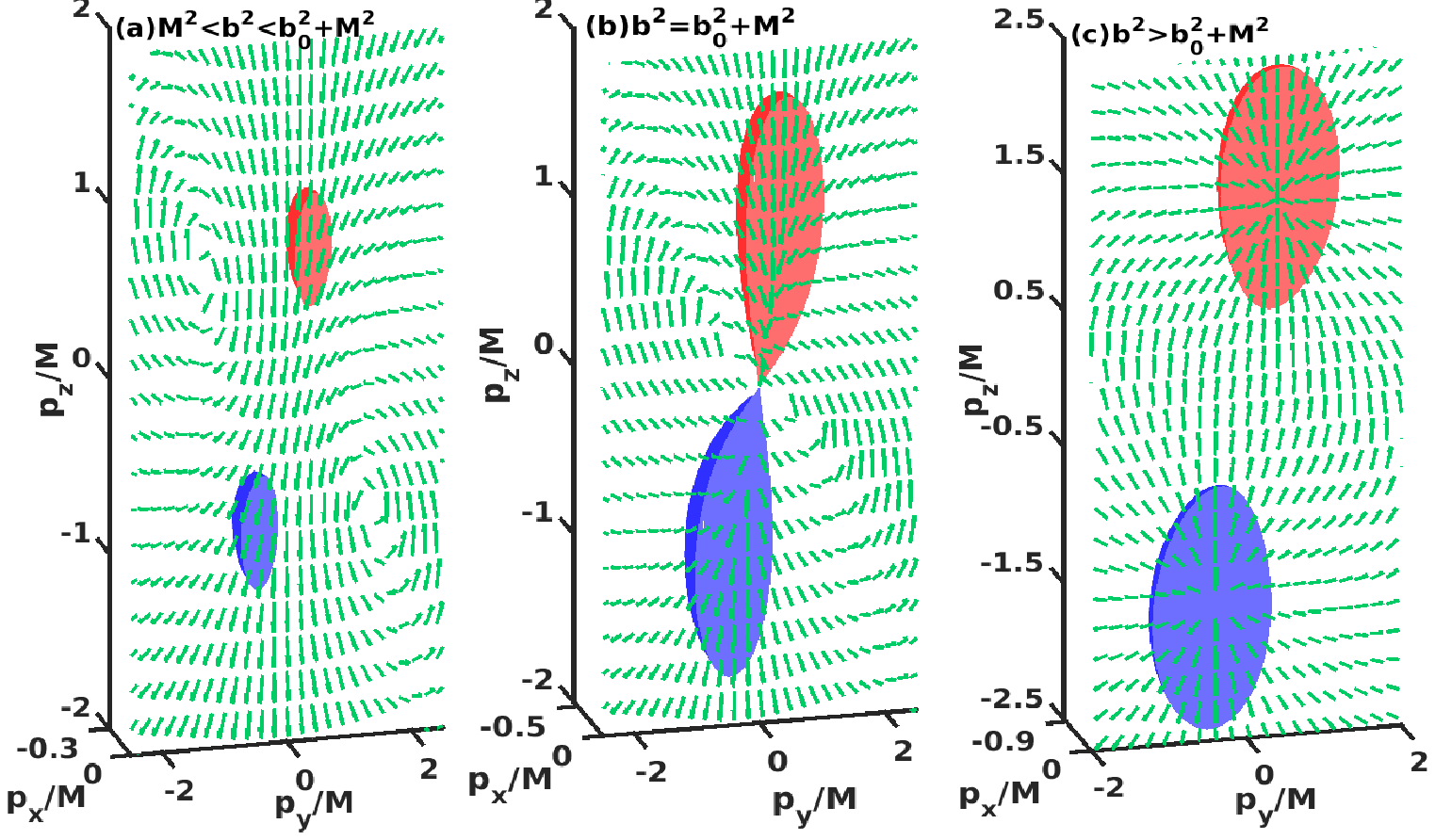} 
\caption{\textbf{Illustration of the process of topological Lifshits transition
via type-II marginal Dirac point.} Relativistic Weyl fermions in Eq.(\ref{eq:Fermipoint})
are considered. \cite{KlinkhamerVolovik2005a,Volovik2007} 
\\
(\textit{right}):
Fermi surfaces of the right and left Weyl fermions. They enclose the
Berry phase monopoles
with topological charge $N_{3}=+1$ and $N_{3}=-1$ respectively. The monopole configuration is depicted by of green vectors.
\\
(\textit{left}): Fermi surfaces are topologically trivial.
\\
 (\textit{middle}): At the border
between the two regimes the Fermi surfaces are attached to the marginal
Dirac point, which is formed by merging of two Weyl points. 
}
\label{TypeIIDirac} 
\end{figure}


As we mentioned in Sec. \ref{sec:Introduction}, with the multiplicity of topological
invariants for the manifolds of nodes in the fermionic spectrum, Lifshitz transitions become diverse and complex. This can be seen on examples of the Lifshitz transitions with the reconstruction of the Fermi surfaces, where several topological invariants may interplay. 
In Sec. \ref{I-II-transition} and Sec. \ref{EventHorizon} we discussed how the Dirac lines and Fermi surfaces emerge in the Lifshitz transition between two types of the Weyl point. Here we discuss the opposite case, when the type-II Weyl and type-II Dirac points emerge during the Fermi surface Lifshitz transitions. By the  type-II Weyl point the topological invariant $N_3$ is transported
between the Fermi surfaces.

 In general, topological invariants which are involved in the complex topological Lifsihitz
transitions are: (i) the invariant $N_{1}$, which is responsible for
the local stability of the Fermi surface;\cite{Volovik2003}
(ii) the invariant $N_{3}$, which is the global invariant describing the
closed Fermi surface: when the Fermi surfaces collapse to a point, it becomes the type-I Weyl point
with the topological charge $N_{3}$; and 
(iii) the $N_{2}$ invariant in Eq.(\ref{eq:N2}) 
which characterizes the Dirac line.
All three topological invariants are involved in the complex Lifshitz transition.
 For Fermi surfaces with non-vanishing $N_{3}$, there is the type-II point attached to the Fermi surfaces
at the critical point of Lifshitz transition.  This type-II point 
has also the nontrivial $N_{2}$, with the contour $C$ chosen as the  infinitesimal loop around the cone, see also reference
{[}22{]}.\cite{YongXu2015} This is the consequence of the $\pi$ Berry phase along the  infinitesimal loop 
around the Weyl point. And of course, the invariant $N_1$ supports the local stability of the Fermi surface and does not allow to make a hole in the Fermi surface and disrupt it.

Here, we present three models, each with its own characterstics, which
exhibit complex topological Lifshitz transition induced by the interplay between
$N_{1}$, $N_{3}$ and $N_{2}$ invariants. 

\subsection{Lifshitz transitions via marginal Dirac point}
\label{Sec:intermediateDirac}

Fig. \ref{TypeIIDirac} demonstrates the Lifshitz transition, where
the intermediate state represents the type-II Dirac point. Such transition
has been discussed in relativistic theory with the CPT-violating perturbation.\cite{KlinkhamerVolovik2005a,Volovik2007}
The corresponding Hamiltonian for a massive Dirac particle with mass
$M$ has the form: 
\begin{equation}
H=\begin{pmatrix} {\mbox{\boldmath$\sigma$}} \cdot ( c\,{\bf p}-{\bf b}) - b_0  &M
 \\ M & -{\mbox{\boldmath$\sigma$}} \cdot ( c\,{\bf p}+{\bf b})  +b_0 \end{pmatrix} \,.
\label{eq:Fermipoint}
\end{equation}
Here the 4-vector $b_{\mu}=(b_{0},{\bf b})$ causes the shift of the positions of the
Berry phase monopoles in opposite direction and formation of two Fermi
surfaces with the global charge $N_{3}=\pm1$ if ${\bf b}^{2}>b_{0}^{2}+M^{2}$ as is
shown in Fig.\ref{TypeIIDirac}(c). One Fermi surface enclosing the
Berry phase monopole with topological charge $N_{3}=+1$ is formed
by the right-handed Weyl fermions; while the other one, which encloses
the Berry phase monopole with topological charge $N_{3}=-1$, comes
from the left-handed Weyl fermions. Positions of the monopoles
are at
\begin{equation}
{\bf p}_\pm = \pm{\bf b} \left(\frac{{\bf b}^2-b_0^2-M^2}{{\bf b}^2-b_0^2}\right)^{1/2} \,.
\label{Monopoles}
\end{equation}

At critical point of Lifshitz transition,
${\bf b}^{2}=b_{0}^{2}+M^{2}$,  two Berry phase
monopoles with opposite chirality merge forming the Dirac point with trivial topological charge
$N_{3}=0$. In contrast, the non-vanishing $N_{1}$
locally preserves Fermi surfaces at this critical
point. As a result of this interplay between $N_{3}$ and $N_{1}$,
the Fermi surfaces are attached to the Dirac point forming the type-II 
Dirac point, see Fig. \ref{TypeIIDirac}(b).
Note that as different from the Weyl point, the Dirac point  is marginal: it has  trivial global topology and is not stable, if there is no special symmetry
which can stabilize the node.\cite{Volovik2003}
Here  the type-II Dirac point appears exactly at the Lifshitz transition, similar to the 
appearance of the Dirac nodal line in Fig. \ref{3HeA}.

\begin{figure}
\centering 
\includegraphics{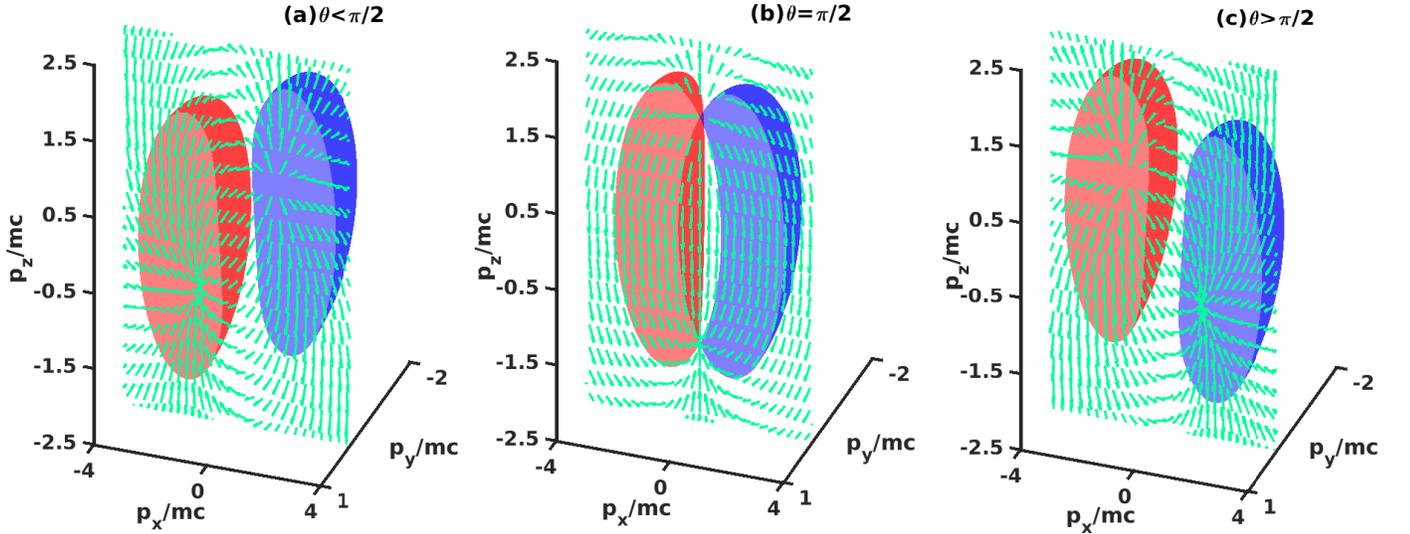} \caption{\textbf{Lifshitz transition and the transport of topological charge
between two Fermi surfaces through the type-II Weyl point in chiral
superlfuid.} (a) $\theta<\pi/2$. In this case, the red Fermi surface
and the blue Fermi surface are both globally non-trivial, with $N_{3}=-1$
and $N_{3}=1$ respectively. Green arrows represent the configuration
of corresponding Berry phase monopoles in the $p_{y}=0$ plane. When
$\theta$ is increased to the case of (b) $\theta=\pi/2$, one can
find that both Fermi surfaces are globally trivial. Equivalently,
two Berry phase monopoles are pushed out from Fermi surfaces and form
two type-II Weyl points as the intermediate states of Lifshitz transition.
(c )$\theta>\pi/2$. Two Berry phase monopoles are pulled into the Fermi
surfaces again and the globally non-trivial property of Fermi surfaces
are revived. The only difference is that the $N_{3}$ of every
Fermi surface with $\theta>\pi/2$ is opposite to what it is in case
(a), i.e.  the red blue Fermi surfaces have $N_{3}=+11$
and $N_{3}=-1$ respectively. So, we can find that the topological charges of Fermi surfaces
are transported between the Fermi surfaces via the type-II Weyl points. }
\label{TypeIIWeyl} 
\end{figure}


\subsection{Lifshitz transitions via Weyl points}
\label{Sec:intermediateWeyl}

For the type of transition discussed in Sec. \ref{Sec:intermediateDirac}, the topological index $N_{2}$ is not involved, because the intermediate state is the Dirac point. 
To obtain  the type-II fermions with non-vanishing $N_{2}$ at the Lifshitz
transition, one should consider the system in which 
the Fermi surfaces are connected not by the type-II marginal Dirac point but by the
topologically stable type-II Weyl point.

The corresponding Hamiltonian is obtained by the natural extension of  Eq.(\ref{Hamiltonian3HeA})  for quasiparticles in chiral superfluid $^3$He-A in the presence of the superfluid current. If the current is chosen perpendicular to the directions towards the Weyl nodes, then at $v>c$ we obtain two type-II Weyl points, which connect two banana shape Fermi surfaces in Fig.\ref{TypeIIWeyl} (b). 
These type-II Weyl points, in addition to the Berry monopole invariant $N_3=\pm 1$, have the 
nonzero value of the topological charge
 $|N_{2}|$ calculated for the infinitesimal closed loop $C$ around the cones. If the closed loop $C$
 is within the symmetry plane of two
Fermi surfaces, the integral is independent on the shape and the radius of the contour $C$ of
 integration. In general, however, when the symmetry is violated, only the integration over the infinitesimal 
loop gives the integer value of the invariant, see Sec.\ref{Sec:LoosingMonopole}.

Let us now change the direction of the current.
If $\theta$ is the angle between the current and the directions to the nodes, the Hamiltonian in the laboratory
frame becomes:
\begin{equation}
H=p_{x}v+\tau_{3}\frac{p^{2}-p_{F}^{2}}{2m}+\tau_{1}c(p_{x}\sin\theta-p_{z}\cos\theta)+\tau_{2}cp_{y}\,.\label{TypeIIWeyl1}
\end{equation}
Eq.(\ref{TypeIIWeyl1}) is identical to Eq.(\ref{Hamiltonian3HeA}) when $\theta=\pi/2$.
Fig. \ref{TypeIIWeyl}  demonstrates the Lifshitz
transition induced by the change of $\theta$, from $\theta<\pi/2$
to $\theta>\pi/2$, at $v>c$.  When one continuously
changes the angle $\theta$ across $\theta=\pi/2$, the Berry phase monopoles
with topological charges $N_3=\pm 1$   
move counter clockwise on a Fermi sphere with radius $|\mathbf{p}|=p_{F}$
in the $p_{y}=0$ plane.   At the same time, the non-vanishing local stability
invariant $N_{1}$ with $v>c$ protects the Fermi surfaces during this process.
As a result, the Berry phase monopoles are transported between the two
Fermi surfaces, with the type-II Weyl points emerging in the intermediate state of
this topological Lifshitz transition.
Similar phenomenon with the interplay between topological invariants
$N_{1}$, $N_{2}$  and $N_{3}$ may take place in $bbc$ Fe,
see details in Ref. \onlinecite{GosalbezMartinez2015}.

\begin{figure}[htb]
\includegraphics[scale=0.6]{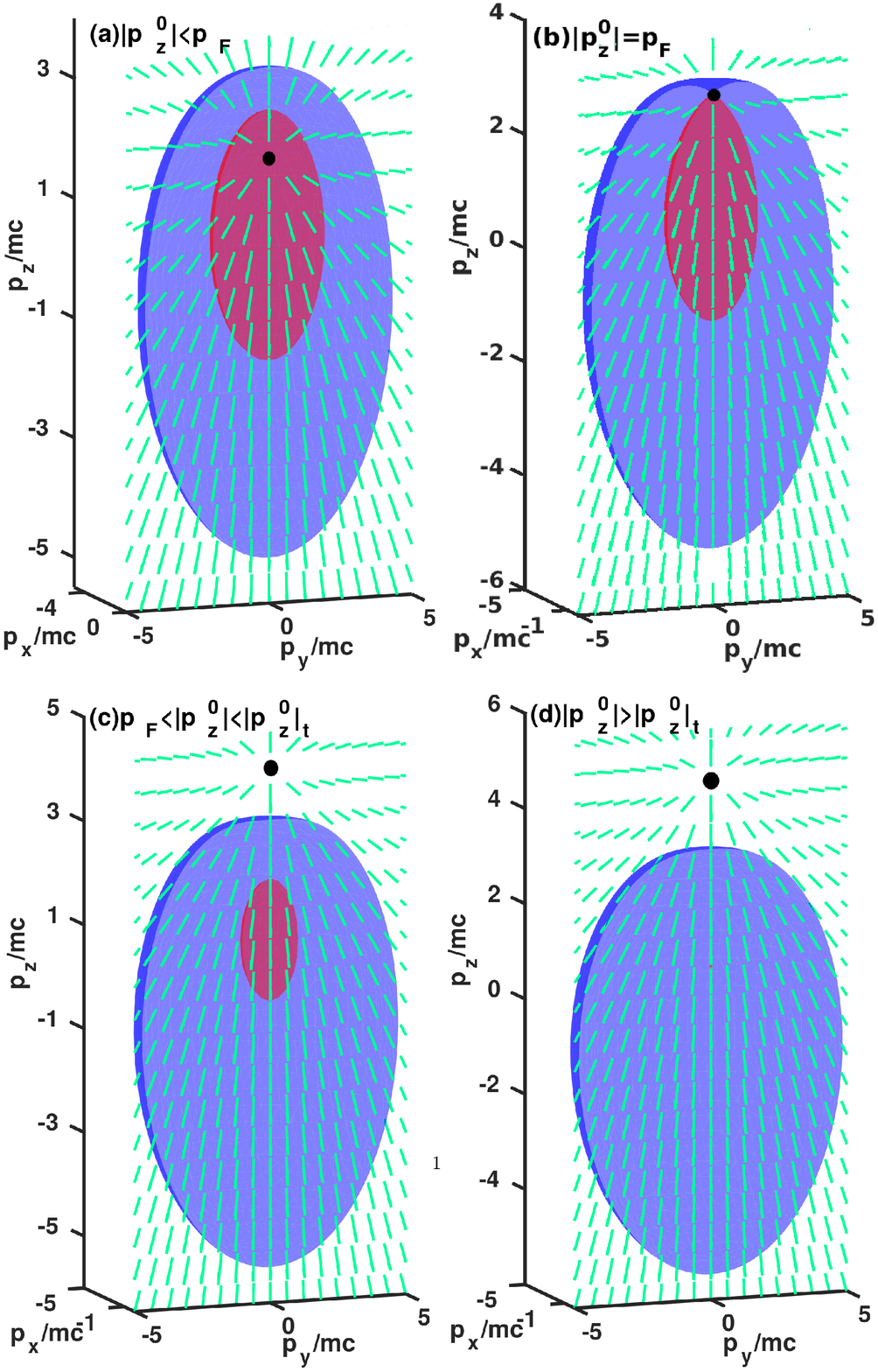}
\caption{\textbf{Illustration of the process in which Fermi surfaces loose their global topological charge.}
\\
  (\textit{top left}): 
Both blue and red Fermi surfaces enclose the Berry  monopole with topological
charge $N_{3}=+1$, when  $|\mathbf{p}^{(0)}| < p_F$. 
\\
(\textit{top right}): The intermediate state of Lifshitz transition at  $|\mathbf{p}^{(0)}| = p_F$. The inner and outer Fermi surfaces touch each other,
at  the peculiar type-II Weyl point with topological charge $N_{3}=+1$.  
\\
(\textit{bottom left}):  On the other side of the Lifshitz transition, at  at  $|\mathbf{p}^{(0)}| > p_F$, the Weyl point is outside of the Fermi surfaces, i.e. after the transition the Fermi surfaces lost the Berry phase flux. 
\\
 (\textit{bottom right}): After the second Lifshitz transition, which takes place at  at $|\mathbf{p}^{(0)}| =(m^2c^2 + p_F^2)/2mc$, the inner Fermi surface disappears, since it is not protected by the topological charge $N_3$.
 }
\label{TypeIIWeyl2} 
\end{figure}


\subsection{Fermi surface  looses Berry monopole after Lifshitz transition}
\label{Sec:LoosingMonopole}

Let us consider another class of emergent type-II Weyl point, in which the Berry
phase monopole is transported across the Fermi surface. It can be represented by the following
Hamiltonian:
\begin{equation}
H=c{\mbox{\boldmath\ensuremath{\sigma}}}\cdot({\bf p}-{\bf p}^{(0)})+\frac{p^{2}-p_{F}^{2}}{2m}\,
\label{ExampleHamiltonian}
\end{equation}

In Fig. \ref{TypeIIWeyl2}, we plot the Lifshithz transitions and the
evolution of configuration of Berry  monopole in momentum space
driven by the change of the position $\mathbf{p}^{(0)}$ of the Weyl point. The regime with $p_F > mc$ is considered.
For $|\mathbf{p}^{(0)}| < p_F$ we have two Fermi surfaces, one inside the other, but both embracing
the Weyl point with $N_3=1$, the Berry phase monopole. At the Lifshitz transition, which occurs  at  
$|\mathbf{p}^{(0)}| =p_F$, the inner and outer Fermi surfaces touch each other at the Weyl point, which becomes the peculiar type-II point. As distinct from the conventional type-II Weyl point, which connects two Fermi pockets,  this Weyl point connects the inner and outer Fermi surfaces. After the Lifshitz transition, at  $|\mathbf{p}^{(0)}| >p_F$, the Weyl point leaves both Fermi surfaces. The Fermi surfaces are again one inside the other, but 
both without the Berry flux. Finally at the second Lifshitz transition, at $|\mathbf{p}^{(0)}| =(m^2c^2 + p_F^2)/2mc$,  the inner Fermi surface collapses to the point and disappears, since the point is no more supported by the topological invariant $N_3$.

At the first Lifshitz transition, the cone formed at the touching point is again characterized by the topological invariant $N_2=1$, where 
the integral is over the infinitesimal path around the cone.

\section{Flat bands at Lifshitz transitions}
\label{Sec:FlatBand}

For the interacting fermions more types of Lifshitz transitions are possible -- the transitions  which involve the Weyl points of type-III and type-IV.\cite{NissinenVolovik2017} 
The interaction also leads to the formation of the flat band in the
energy spectrum --  the so-called Khodel-Shaginyan fermion 
condensate.\cite{KhodelShaginyan1990,Volovik1991,Nozieres1992}
The dispersionless energy spectrum has a singular density of states. As a results,  instead of the exponential suppression of the superconducting transition
temperature $T_c$ (and of the gap $\Delta$)  in the normal metal, the flat band provides $T_c$ and $\Delta$ being proportional to the coupling constant $g$ in the Cooper channel:
\begin{equation}
\Delta_{\rm normal} = E_0  \exp\left(- \frac{1}{ g N_F}\right) \,\,\,,
 \,\,\Delta_{\rm flat\, band}  = \frac{ gV_d}{2(2\pi \hbar)^d}\,.
\label{expVSlin}
\end{equation}
Here $N_F$ is the density of states in normal metal; $d$ is the dimension of the metal; and $V_d$ is the volume of the flat band. For  nuclear systems, i.e. for $d=0$,  
the linear dependence of the gap on the coupling constant has been found by Belyaev.\cite{Belyaev1961}
The enhancement of $T_c$  in materials with the flat band opens the route to room temperature superconductivity, see review Ref.\cite{HeikkilaVolovik2016}.

The band flattening caused by electron-electron interaction in metals is the manifestation of the general phenomenon of the energy level merging due to electron-electron interaction. This effect has been recently
suggested \cite{Dolgopolov2014} to be responsible for merging  of the discrete energy levels in two-dimensional electron system in quantizing magnetic fields. 

 According to Ref.~\cite{Yudin2014} the favourable condition for the formation of such flat band is when the van Hove singularity comes close to the Fermi surface, i.e. the system is close to the Lifshitz transition  (see also Ref.~\cite{Volovik1994} for the simple Landau type model of the formation of such flat band). It is also possible that this effect is responsible for the occurrence of superconductivity with high $T$ observed in the pressurized sulfur hydride \cite{Drozdov2014,Drozdov2015}.
There are some theoretical evidences \cite{Pickett2015,Bianconi2015}  that the high-$T_c$. superconductivity takes place at such pressure, when the system is near the Lifshitz transition
That is why it is not excluded that  the Khodel-Shaginyan flat band is formed  in sulfur hydride at pressure 180-200 GPa giving rise to high-T$_c$ superconductivity. The topological Lifshitz transitions with participation of the Weyl and Dirac points and Dirac lines may also lead to the formation of the flat bands in the vicinity of transitions, and thus to the enhanced $T_c$.

\begin{figure}[ht]
     \begin{center}        
   \includegraphics[width=0.78\textwidth,]{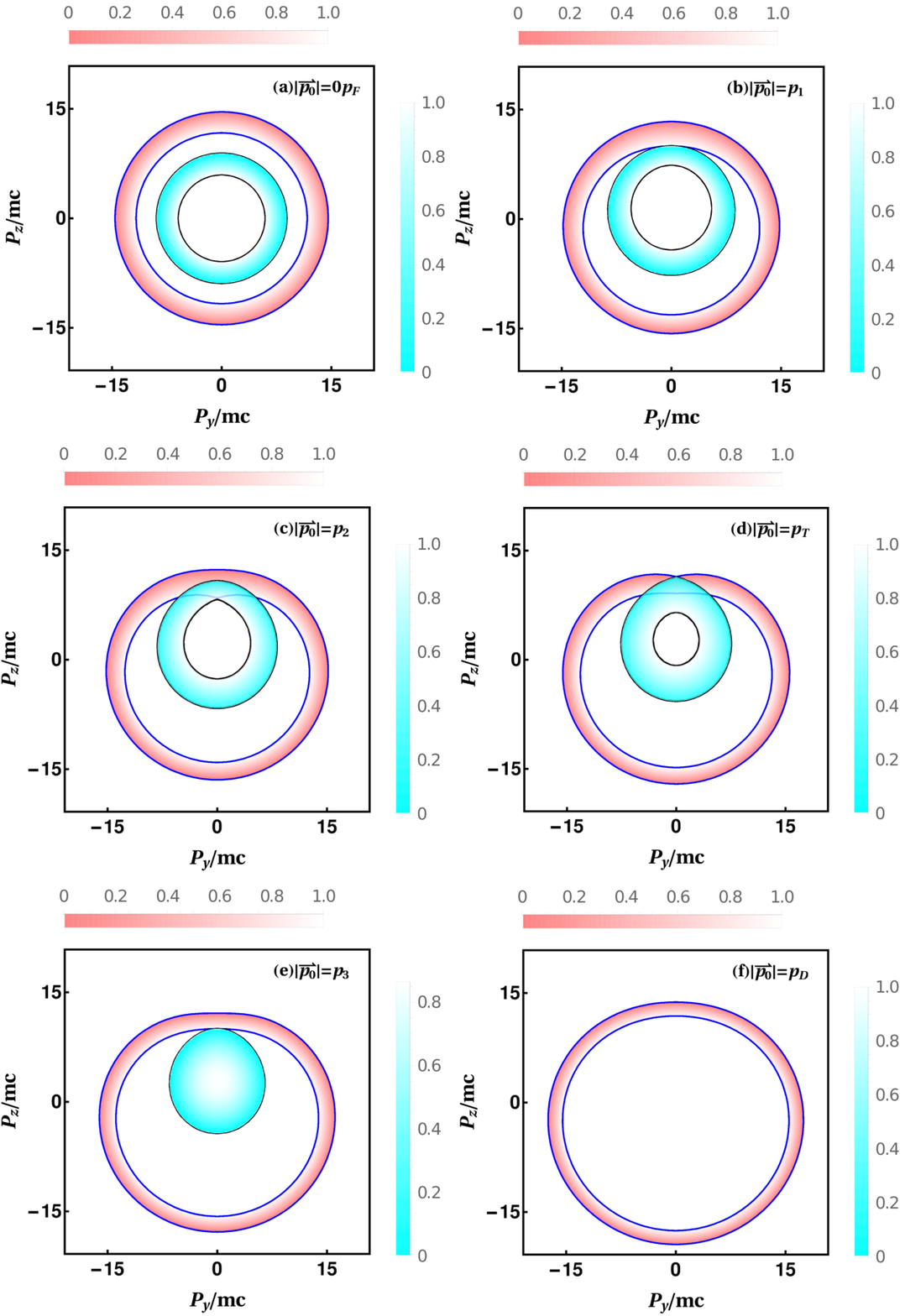}\\ 
            
    \end{center}
    \caption{Flat band of loosing monopole Hamiltonian (\ref{ExampleHamiltonian}) as a function of parameter  $|\mathbf{p}^{0}|$ for $U=0.4E_{F}$, $U_{m}=0.384E_{F}$, $p_{F}=10mc$ and $E_{F}=50mc^{2}$ in $p_{y}-p_{z}$ plane. The red flat band and the blue flat band correspond to distribution functions of particle and hole respectively. The color bars depict the occupation probability of particle or hole. While in the non-interacting case two Fermi surfaces touch each other at $|\mathbf{p}^{0}|=p_{F}$, the touching of the Fermi bands occurs earlier, at
 $|p^{0}|=p_{1}<  p_{F}$, see (b), with $p_1$  from Eq.(\ref{1T}). Above $p_{1}$, these two flat bands overlap. Another  transition happens  at $|\mathbf{p_{0}}|=p_2$ in (c)  with $p_{2}$ from Eq.({\ref{2T}}), 
  where the band touching looks similar to the touching of the Fermi surfaces at the Weyl type-II point.
 The similar transition takes place
 at $|\mathbf{p_{0}}|=p_{T}$ in (d) with $p_{T}$ from Eq.({\ref{HT}}), where the band touching looks similar to the touching of the Fermi surfaces at the Weyl type-II point. This is the common point at which the occupation probabilities vanish both for particle and hole. At $|\mathbf{p}^{0}|=p_{3}$ in Eq.(\ref{3T}), the inner band contacts the outside band at one point on the symmetry axis,  as shown in (e). The inner flat band finally disappears at $|\mathbf{p}^{0}|=p_{D}$ with $p_{D}$ in Eq.(\ref{D}), as shown in (f).  } %
   \label{FBLW}
\end{figure}

Here we consider the flat bands, which appear near the Lifshitz transition  
in Eq. (\ref{ExampleHamiltonian}). The arrangement of the flat bands experiences its own Lifshitz transitions in Fig. \ref{FBLW}.

The energy functional of interacting loosing Weyl model is:
\begin{equation}
E[n_{1}(\mathbf{p}),n_{2}(\mathbf{p})]=\sum _{\mathbf{p}}{{\epsilon}_{1 }^{0}{n}_{1}(\mathbf{p})
+{\epsilon}_{2}^{0}{n}_{2}(\mathbf{p})
+\frac{1}{2} U(n_{1}(\mathbf{p})-\frac{1}{2})^{2}
+\frac{1}{2} U(n_{2}(\mathbf{p})-\frac{1}{2})^{2}
+\frac{1}{2} U_{m}({n}_{1}(\mathbf{p})-\frac{1}{2})({n}_{2}(\mathbf{p})-\frac{1}{2})}\,,
\end{equation} 
where $n_{1}(\mathbf{p})$ and $n_{2}(\mathbf{p})]$ are distribution functions for two species of fermions ("partilces" and "holes"). For the flat bands induced by the repulsive interaction, we have:
\begin{equation}
{\epsilon}_{1}=\frac{{\delta}E}{{\delta} n_{1}(\mathbf{p})}=0 \,\,,\,\,{\epsilon}_{2}=\frac{{\delta}E}{{\delta} n_{2}(\mathbf{p})}=0\,,
\label{e12}
\end{equation}
which give
\begin{eqnarray}
{\epsilon}_{1}^{0}+U(n_{1}(\mathbf{p})-\frac{1}{2})+\frac{1}{2}U_{m}(n_{2}(\mathbf{p})-\frac{1}{2})=0\,,
\\
{\epsilon}_{2}^{0}+U(n_{2}(\mathbf{p})-\frac{1}{2})+\frac{1}{2}U_{m}(n_{1}(\mathbf{p})-\frac{1}{2})=0\,.
\end{eqnarray}
The distribution functions of particles and hole are:
\begin{equation}
n_{1}(\mathbf{p})=\frac{
4U^{2}-U_{m}^{2}-8U{\epsilon}_{1}^{0}+4U_{m}{\epsilon}_{2}^{0}}{8U^{2}-2U_{m}^{2}},
\end{equation}
and
\begin{equation}
n_{2}(\mathbf{p})=\frac{
4U^{2}-U_{m}^{2}-8U{\epsilon}_{2}^{0}+4U_{m}{\epsilon}_{1}^{0}}{8U^{2}-2U_{m}^{2}}.
\end{equation}
Those regions within which $0<n_{1}(\mathbf{p})<1$ and $0<n_{2}(\mathbf{p})<1$ are the flat bands in momentum space, as shown in Fig.(\ref{FBLW}). With increasing $|\mathbf{p}^{0}|$, several Lifshitz transitions occur at the critical points of Lifshitz given by:
\begin{equation}
p_{1}=p_{F}-\frac{2U-U_{m}}{4c},
\label{1T}
\end{equation}
\begin{equation}
p_{2}=\sqrt{p_{F}^{2}-\frac{2mU+mU_{m}}{2}},
\label{2T}
\end{equation}
\begin{equation}
p_{T}=\sqrt{p_{F}^{2}+\frac{2mU+mU_{m}}{2}},
\label{HT}
\end{equation}
\begin{equation}
p_{3}=p_{F}+\frac{2U-U_{m}}{4c},
\label{3T}
\end{equation}
\begin{equation}
p_{D}=\frac{2c^{2}m^{2}(2U+U_{m})^{2}+(U_{m}-2U)^{2}[2p_{F}^{2}+m(2U+U_{m})]}{4mc(4U^{2}-U_{m}^{2})}.
\label{D}
\end{equation}
At these transitions the Fermi bands appear, disappear or touch each other with formation of singular configurations.

\section{Conclusion}

The interplay of different topological invariants enhances the variety
of the topological Lifshitz transitons. Here we discussed the examples
of the transitions, which involve the nodes of different co-dimensions: the Fermi surfaces with topological
charge $N_{1}$ (co-dimension 1), Weyl points with the topological charge $N_{3}$ (co-dimension 3)
and Dirac lines with topological charge $N_{2}$ (co-dimension 2). Depending on the
type of the transition, the intermediate state has the type-II Dirac
point, the type-II Weyl point or the Dirac line. The latter is supported by combination of symmetry
and topology. There are different configurations of the Fermi surfaces, involved in the  Lifshitz transition with the Weyl points in the intermediate state.
In Fig. \ref{TypeIIWeyl} the type-II Weyl point connects
the Fermi pockets, and the Lifshitz transition corresponds to the transfer of the Berry flux between the Fermi pockets.
In Fig.\ref{TypeIIWeyl2} the type-II Weyl point connects the outer and inner Fermi surfaces. At the Lifshitz transition the Weyl point is released from both Fermi surfaces. They loose their Berry flux and the 
topological charge $N_3$, which guarantees the global stability. As a result the inner surface disappears after shrinking to a point at the second Lifshitz transition.

Many other Lifshitz transitions are expected, since
we did not touch here the other possible topological features: topological
invariants which describe the shape of the Fermi surface; the shape
of the Dirac nodal lines; their interconnections; etc. The interplay
of topologies can be seen in particular in the electronic spectrum
of Bernal and rombohedral graphite.\cite{mcclure57,Mikitik2006,Mikitik2008,HeikkilaVolovik2015}
In particular, in the electronic spectrum of Bernal graphite the type-II
Dirac line has been identified, which is connected with the type-I
Dirac line at some point in the 3D momentum space.\cite{HyartHeikkilaVolovik2016}
If one considers $p_{z}$ as parameter, then at some critical value
of $p_{z}$ there is the transition from the 2D type-I Dirac point
to the 2D type-II Dirac point.

However, the most important property of Lifshitz transitions is that in the vicinity of the topological transtion the electron-electron interaction leads to the formation of zeroes in the spectrum of the co-dimension 0, i.e. to the flat bands. Because of the singular density of electronic states, materials with the flat band are the plausible candidates 
for room-temperature superconductivity.

\section{Acknowledgements}
We thank Ivo Souza for pointing out mistake in the early version, and
Tero Heikkil\"a for discussion on the type-II Dirac lines. The work by GEV 
has been supported by the European Research Council
(ERC) under the European Union's Horizon 2020 research and innovation programme (Grant Agreement No. 694248). The work by KZ has been supported in part by the National Natural Science Foundation of China (NSFC) 
under 
Grants No. 11674200, No. 11422433, No. 11604392.

\end{document}